\documentclass[conference]{IEEEtran}
\ifCLASSINFOpdf
\else
\fi
\usepackage{balance}
\usepackage{graphicx}
\usepackage{subfigure}
\usepackage{amsmath}
\usepackage{amssymb}
\usepackage{color}
\usepackage{paralist}
\usepackage{comment}
\usepackage{cite}
\usepackage{url}
\usepackage{algorithm}
\usepackage{algorithmic}
\usepackage{bibspacing}
\usepackage{hyperref}
\usepackage{empheq}
\usepackage{gensymb}
\usepackage{multirow}
\usepackage{hhline}
\newtheorem{assumption}{Assumption}

\begin{document}

\title{Data-Driven Evaluation of Building\\Demand Response Capacity}

\author{
	\IEEEauthorblockN{Deokwoo Jung\IEEEauthorrefmark{1}, Varun Badrinath Krishna\IEEEauthorrefmark{1}, William G. Temple\IEEEauthorrefmark{1}, David K. Y. Yau\IEEEauthorrefmark{1}\IEEEauthorrefmark{2}} 
	    \IEEEauthorblockA{\IEEEauthorrefmark{1}Advanced Digital Sciences Center, Singapore
	    \\\{deokwoo.jung, varun.bk, william.t, david.yau\}@adsc.com.sg}
	    \IEEEauthorblockA{\IEEEauthorrefmark{2}Singapore University of Technology and Design, Singapore
	    \\\{david\textunderscore yau\}@sutd.com.sg}
	}
    
\maketitle

\begin{abstract}
Before a building can participate in a demand response program, its facility managers must characterize the site's ability to reduce load. Today, this is often done through manual audit processes and prototypical control strategies. In this paper, we propose a new approach to estimate a building's demand response capacity using detailed data from various sensors installed in a building. We derive a formula for a probabilistic measure that characterizes various trade-offs between the available demand response capacity and the confidence level associated with that curtailment under the constraints of building occupant comfort level (or utility). Then, we develop a data-driven framework to associate observed or projected building energy consumption with a particular set of rules learned from a large sensor dataset.
We apply this methodology using testbeds in two buildings in Singapore: a unique net-zero energy building and a modern commercial office building. 
Our experimental results identify key control parameters and provide insight into the available demand response strategies at each site. \footnote{\copyright 2014 IEEE. Personal use of this material is permitted. Permission from IEEE must be obtained for all other uses, in any current or future media, including reprinting/republishing this material for advertising or promotional purposes, creating new collective works, for resale or redistribution to servers or lists, or reuse of any copyrighted component of this work in other works.
Available on \href{http://ieeexplore.ieee.org/xpl/articleDetails.jsp?arnumber=7007703}{IEEE Xplore} DOI: \href{http://dx.doi.org/10.1109/SmartGridComm.2014.7007703}{10.1109/SmartGridComm.2014.7007703}
}
\end{abstract}



%
\IEEEpeerreviewmaketitle
\section{Introduction}
\label{sec:intro}

An efficient demand response (DR) program should take full advantage of the DR potential of each participating consumer. In order to facilitate this, it is crucial to determine the maximum reduction in energy consumption that can be reliably achieved in a building. We refer to this consumption as the \emph{demand response capacity} of the building, and it can be achieved with a certain probability.

To determine the DR capacity of a building, a large number of building-specific models and parameters need to be estimated. It is also necessary to predict the building's energy consumption behavior with respect to various parameter settings. Current approaches for achieving these requirements have several limitations.
One common approach is to apply standardized regression models using power meter data to find a rough estimate of DR capacity. This often fails to capture the unique characteristics of a particular building that determine its DR capacity.
Alternatively, one can use advanced simulation tools to precisely evaluate DR capacity under various parameters settings.
This, however, requires detailed building specifications and in-depth domain knowledge to model buildings with appropriate parameters. Hence, such simulation approaches do not scale well when applied to several buildings.
Furthermore, none of the aforementioned methods can quantify and estimate the DR capacity with respect to a well-defined reliability measure in a way that is consistent across buildings.

In this paper, we introduce a data-driven approach to estimate DR capacity using a probabilistic measure and a rich dataset obtained from two buildings. Our proposed measure captures insightful trade-offs between DR capacity and its associated reliability.
We derive a formula for this measure that is analogous to the Bit Error Rate (BER) formula~\cite{proakis2001}, which is a fundamental performance metric in communication theory. Hence, our work can be potentially extended to more complex scenarios (that may contain multiple DR participants and DR aggregators) employing many useful concepts and mathematical tools developed in communication theory.

We adopt a data-driven approach, using sensor data, 
that makes no assumptions about physical or regression models for the building's power consumption. 
In this approach, we use a look-up table that associates observed or projected consumption with a particular set of conditions (or rules) learned from a large historical data set. Our approach will become increasingly attractive as building data becomes more readily available to building stakeholders through building management systems, sensor devices, and public databases. Those resources allow 
 us to obtain highly granular information about indoor/outdoor conditions (e.g., lighting, air temperature, and humidity), occupant movements (via passive infrared sensors) and electricity consumption (via smart plugs) for a more precise look-up table.

In this paper, we apply our proposed DR capacity measure and data-driven approach 
using historical data from two testbed sites in Singapore: an office in a net-zero energy building~\cite{marszal2011zeb}, and an office in a typical commercial building. Specifically, we demonstrate the usefulness of our method in designing DR strategies by means of a systematic examination of various trade-offs between DR capacity and reliability with respect to several key factors.

The paper is organized as follows. In Section~\ref{sec:related}, we discuss related work.
In Section~\ref{sec:testbeds}, we introduce the testbed sites and the system implementation that was used to drive this study. In Section~\ref{sec:capacity_model}, we present a theoretical framework for estimating DR capacity based on available control actions and occupant comfort constraints. We then apply this approach to assess the DR capacity of the two testbeds in Section~\ref{sec:analysis}. Finally, we conclude in Section~\ref{sec:conclude}.

\section{Context and Related Work}
\label{sec:related}

There are two stages of demand response capacity evaluation for buildings: (i) audit stage~\cite{DRaudit}, (ii) measurement and verification (M\&V) stage~\cite{DRmv}. An audit takes place before the site enrolls in a DR program; its purpose is to identify the amount of responsive load at the site, and the control actions that will be taken in an event. The M\&V stage occurs after a DR event, and allows aggregators or market authorities to confirm that a specified curtailment did indeed occur. In this work, we are concerned with the demand response audit stage.

Demand response audits~\cite{DRaudit} generally involve a high degree of human effort. Within the last decade, researchers have adopted top-down approaches to characterize typical DR actions and audit procedures. For example, the authors of~\cite{mathieu2011quantifying} analyze utility meter data from dozens of buildings and propose methods to help facility managers identify demand response opportunities. Following a similar direction,~\cite{motegi2007} provides a high-level overview of demand response strategies for commercial buildings. Simulation tools have also been developed to allow building owners to estimate their demand response potential based on a variety of site-specific inputs and typical control strategies~\cite{drqat}.

At a more detailed level, a growing body of work has emerged examining specific building loads for demand response purposes. Such efforts often include the use of sensor networks in the built environment. In this area, much attention has been devoted to plug loads or miscellaneous power~\cite{arnold2013plug,Dawson12}, including specific sources of plug load such as laptops~\cite{murthy2012laptop}.
Outside of plug loads, the other dominant building energy end-uses tend to be lighting and heating, ventilation, and air conditioning (HVAC). For both systems, researchers have studied optimal control using inputs from 
sensor networks~\cite{erickson2010,raziei2013optimal,agarwal2011cycling}.

Our work differs from the above efforts in several important ways. Since we use indoor sensor networks rather than utility metering data, we are able to provide a more detailed view of building energy consumption and demand response potential than top-down audit approaches~\cite{mathieu2011quantifying,drqat}. Although several related works (e.g.,~\cite{erickson2010,Dawson12}) also leverage wireless sensor networks, our scope is more holistic in that it includes multiple electricity end-uses (HVAC, lighting, plug load), as well as consideration for occupants' thermal and visual comfort.

\section{Testbed Sites}
\label{sec:testbeds}

For this work, we deployed wireless sensor networks inside two office spaces in Singapore to gather data on occupancy, occupant comfort, and energy usage. These data are used to develop a DR capacity model that incorporates occupant comfort constraints. In this section we provide details about the two testbed sites, as well as the sensor networks deployed and the resulting data that was collected.

\begin{figure}
  \centering
  \subfigure[ZEB testbed: estimated occupancy]{
  \includegraphics[width=0.95\linewidth]{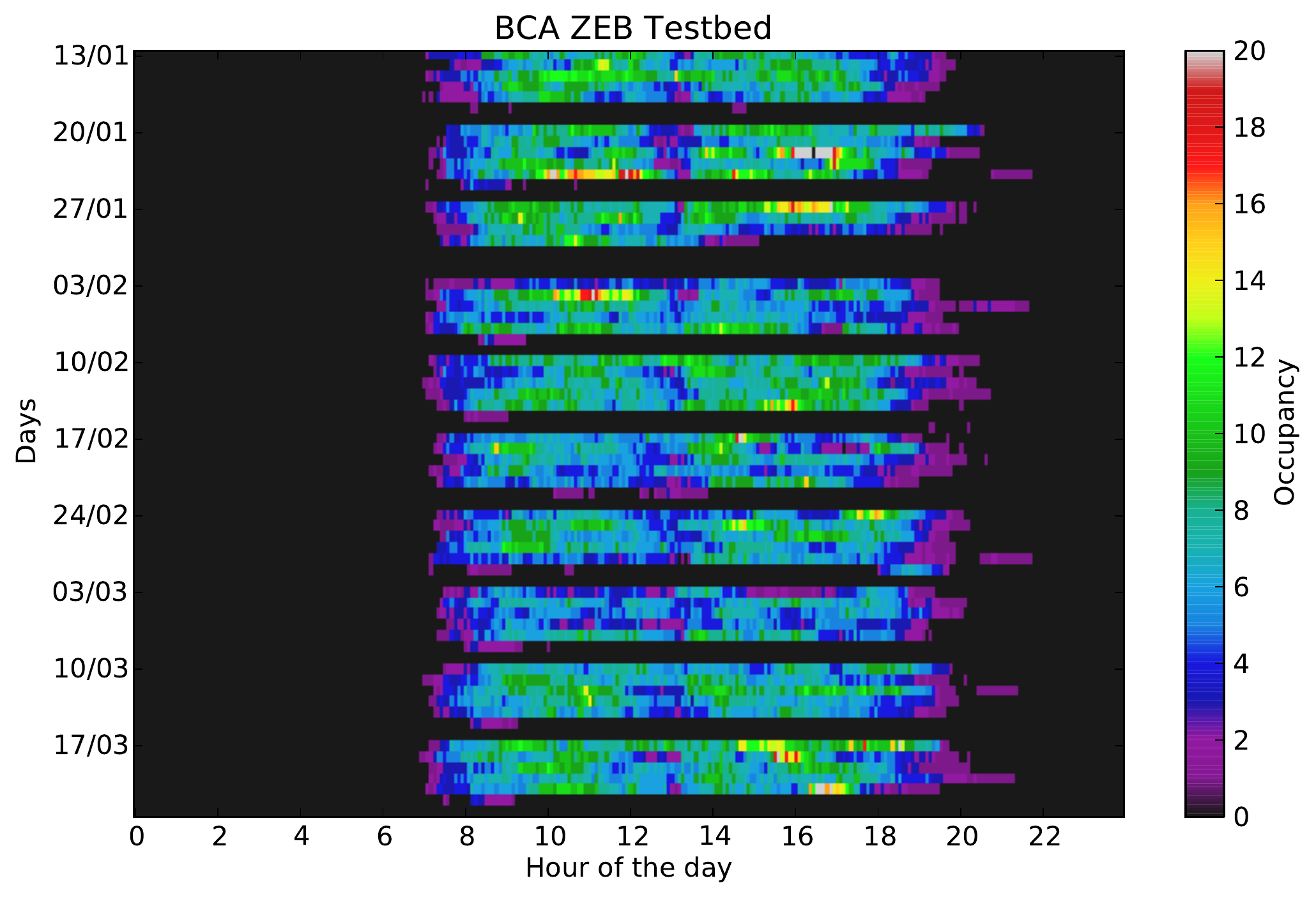}
  }
  \subfigure[ADSC testbed: estimated occupancy]{
  \includegraphics[width=0.95\linewidth]{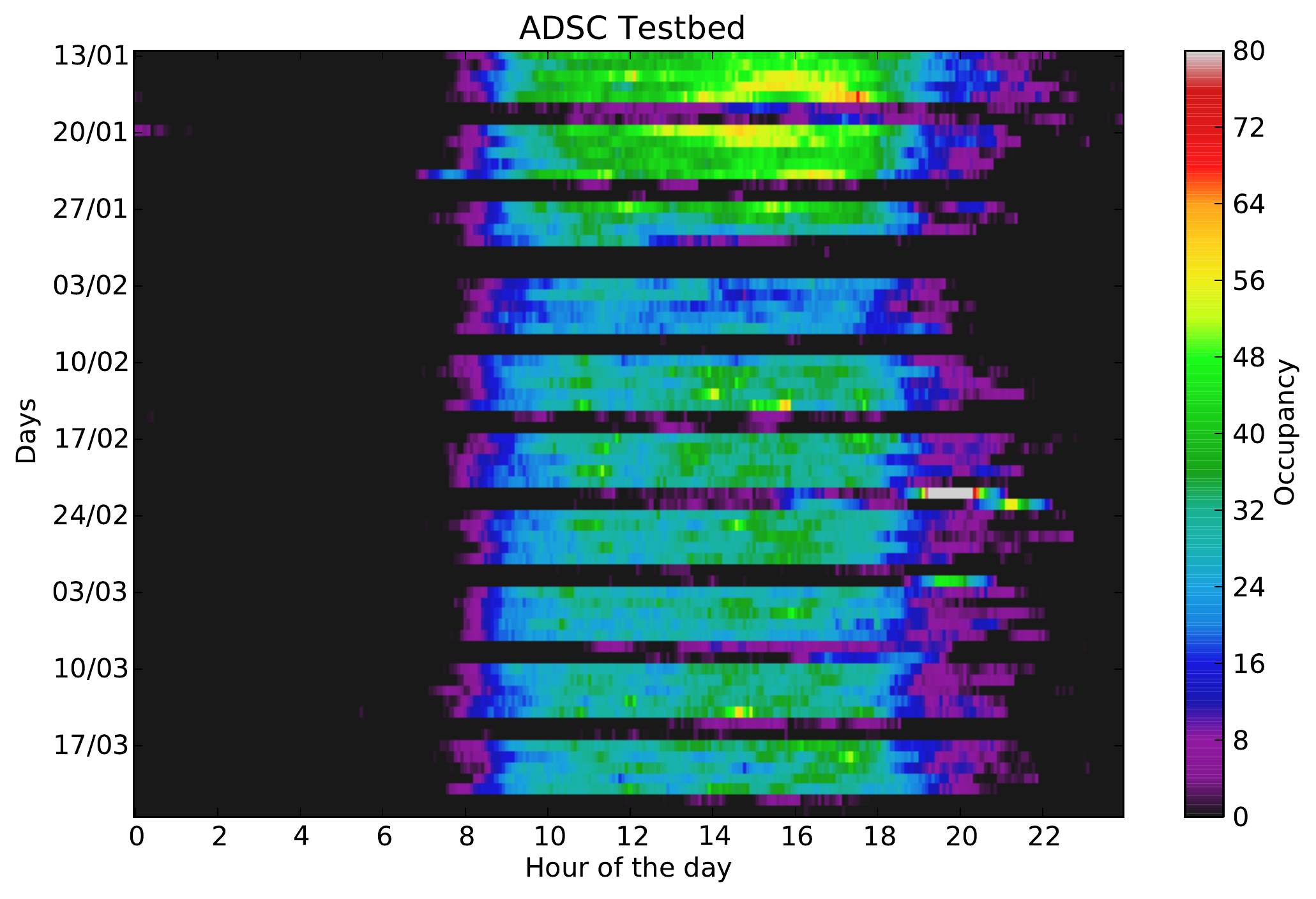}
  }
\caption{Real-time occupancy in two office testbeds over a $10$ week period in $2014$. Occupancy is estimated using the methodology in~\cite{jung2013energytrack}.}
\label{fig:testbeddata-occupancy}
\end{figure}

\subsection{Site Details}
Both testbeds are office spaces, however the features of the buildings are very different.
\textbf{Testbed \#1 (ZEB testbed)} is in a net-zero energy building (ZEB)~\cite{marszal2011zeb} owned by the Building and Construction Authority (BCA) of Singapore. The testbed is located in the office of BCA's Centre for Sustainable Buildings and Construction (CSBC). The ZEB has several energy saving features that are not common in most buildings today, such as light pipes and light shelves (to bring more daylight into the indoor space), as well as displacement ventilation and personalized fresh air supply at individual desks\footnote{\url{https://www.bca.gov.sg/zeb/daylightsystems.html}}. \textbf{Testbed \#2 (ADSC testbed)} is in the authors' office at the Advanced Digital Sciences Center. 
This testbed is typical of a modern commercial office building in Singapore.

In addition to differences in construction, these two sites differ markedly in size, operating hours, and occupancy patterns (see Figure~\ref{fig:testbeddata-occupancy}). 
 The ZEB testbed occupies a $154.5m^2$ area on a single floor of the building, and typically has $10-12$ people working during the day. In contrast the ADSC testbed is much larger (although still on a single floor), at $824.5m^2$, and has a typical occupancy of roughly $40$ people. The working hours at the ADSC testbed are less regular than the ZEB testbed, with the space typically occupied past $8$pm.
In both spaces, personal computers are the primary source of plug load. In the ZEB, however, all employees use laptops rather than desktops, which significantly lowers energy consumption.

\subsection{Sensor Network and Data Collected}
In both testbeds, a wireless sensor network was deployed to carry out real-time occupancy estimation and energy use analysis using the EnergyTrack system~\cite{jung2013energytrack}. The sensor types deployed include temperature/humidity/light (THL) sensors, carbon dioxide ($CO_2$) sensors, passive infrared (PIR) sensors, and smart plug power meters. Figure~\ref{fig:testbeddata-occupancy} contains heat maps showing real-time occupancy estimates for the two testbeds. These estimates are derived from $CO_2$ measurements and PIR trigger events, as described in~\cite{jung2013energytrack}. Knowledge of precise occupancy information is valuable for identifying energy efficiency opportunities, as well as demand response actions that have minimal impact on occupant comfort (see Section~\ref{sec:capacity_model}).

\begin{figure}
  \centering
  \subfigure[ZEB testbed: power consumption from lighting]{
  \includegraphics[width=0.95\linewidth]{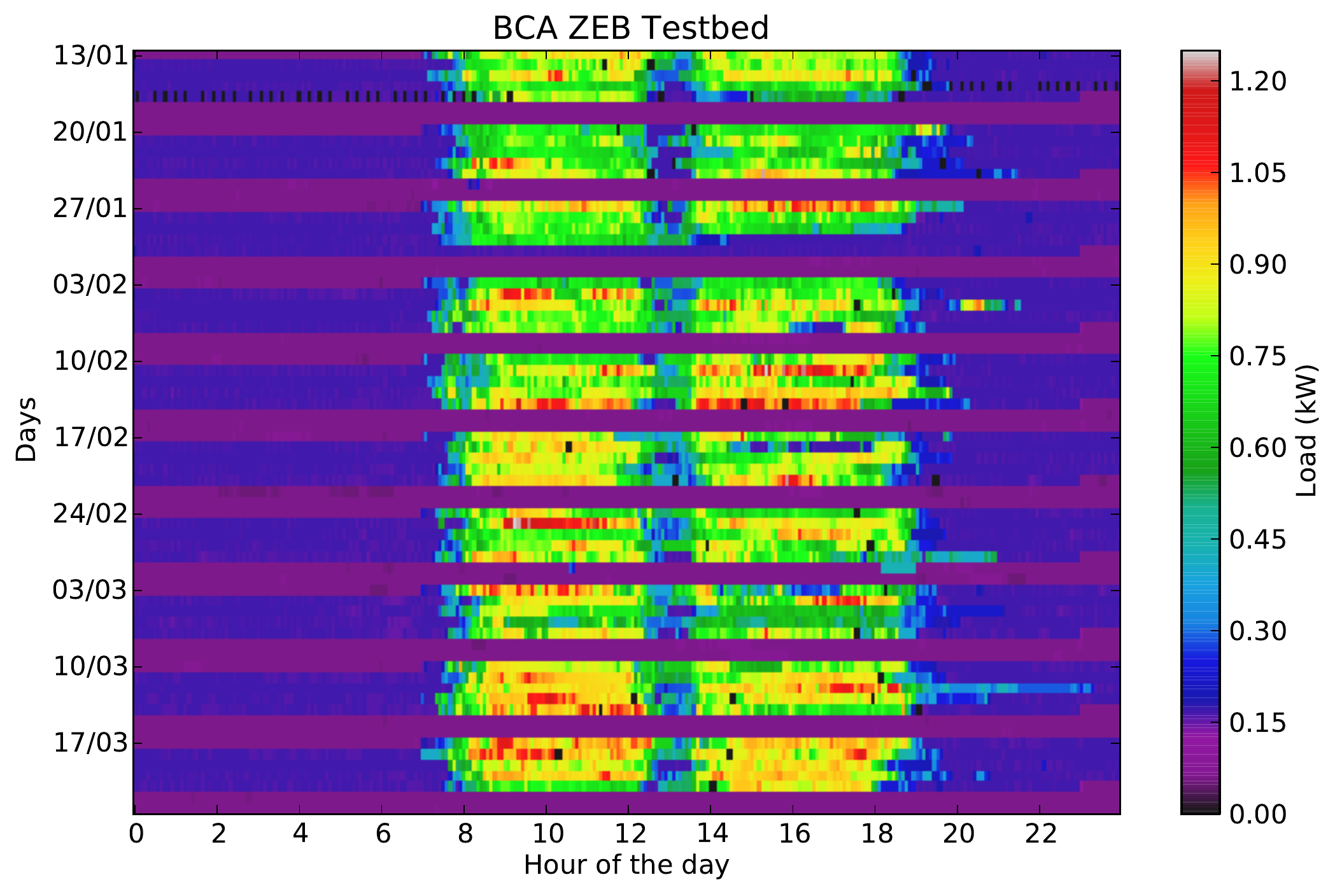}
  }
  \subfigure[ADSC testbed: power consumption from lighting]{
  \includegraphics[width=0.95\linewidth]{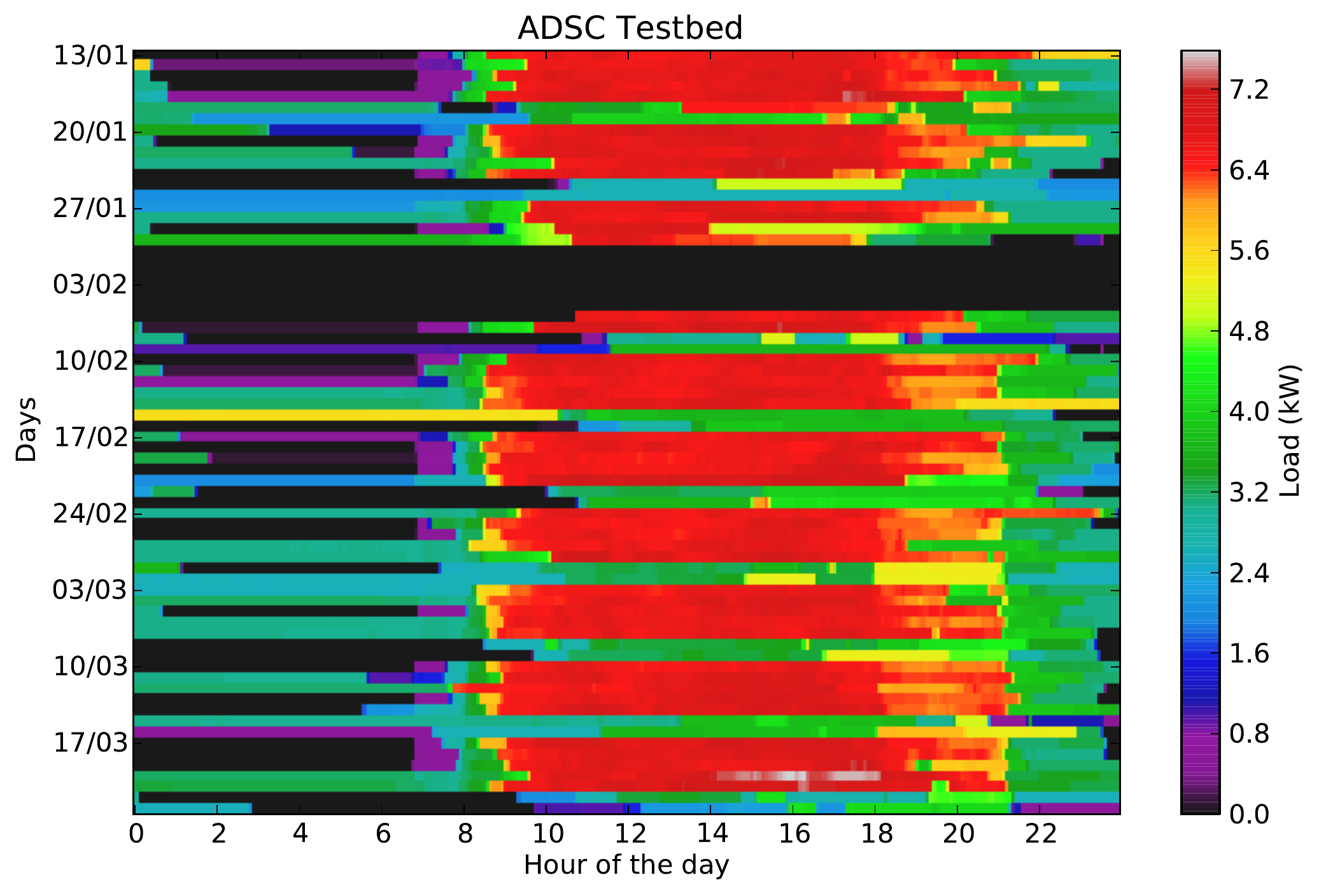}
  }
\caption{Measured lighting power consumption from two office testbeds over a $10$ week period in $2014$.}
\label{fig:testbeddata-light}
\end{figure}

\begin{figure}[t]
  \centering
 \includegraphics[width=0.95\linewidth]{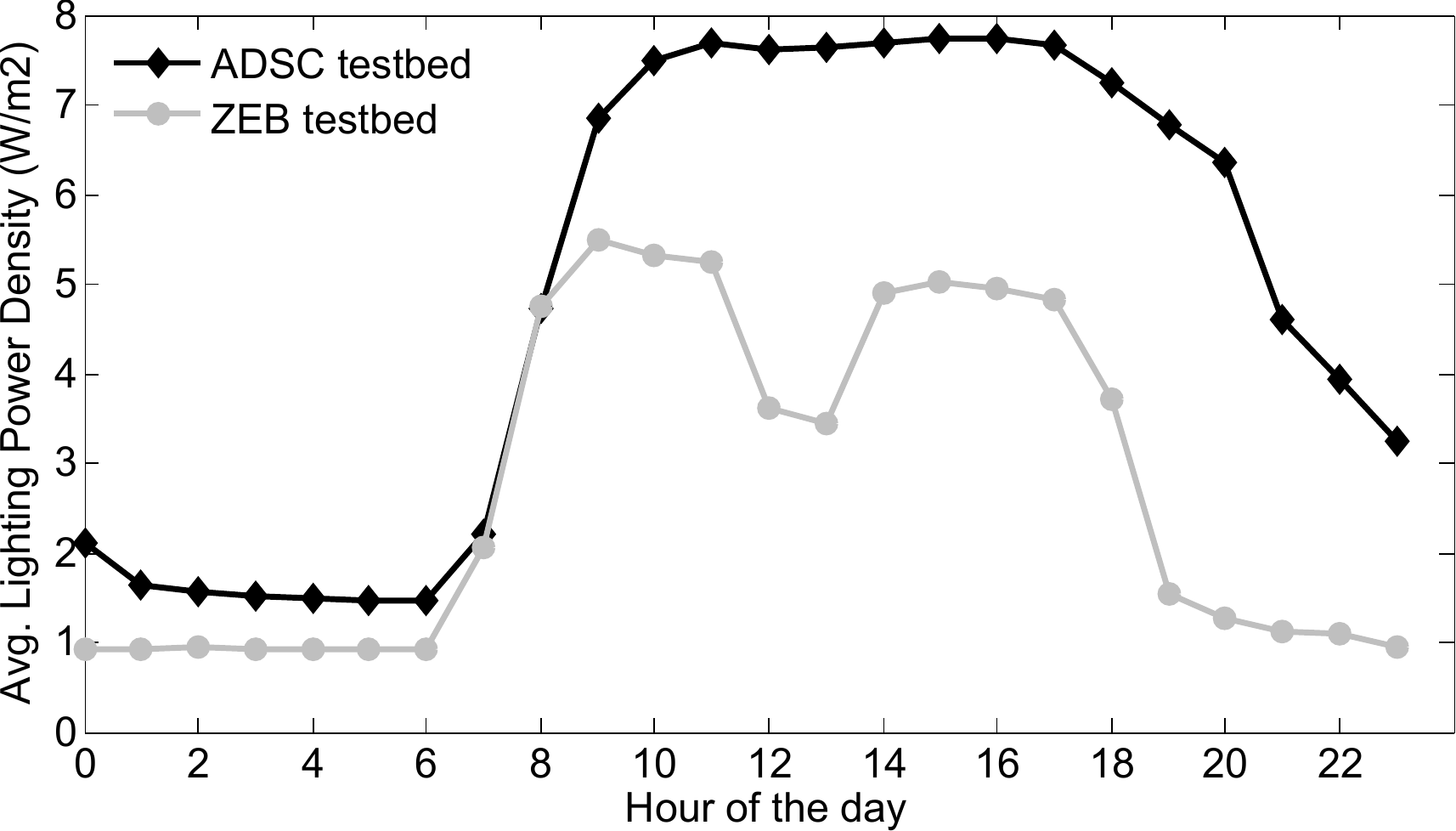}
  \caption{Average lighting power consumption per unit floor area on weekdays, over a $10$ week period in $2014$.}
  \label{fig:avgLightingPwr}
\end{figure}

In the ZEB testbed, lighting and HVAC power consumption data are obtained from the building management system (BMS). In the ADSC testbed, the BMS data were not made available, so lighting power is measured from the electricial distribution panels, and HVAC power consumption is estimated from an EnergyPlus~\cite{energyplus} model. Figure~\ref{fig:testbeddata-light} shows heat maps depicting the measured lighting power consumption in each testbed over the $10$ week data collection period. Similarly, Figure~\ref{fig:avgLightingPwr} shows the average lighting power per unit floor area for weekdays in each of the testbeds. The impact of daylight and dimmable lighting in the ZEB testbed is apparent from the significantly lower lighting power density. 



\section{Measure of demand response capacity} \label{sec:capacity_model}
\label{sec:capacity}
In this section, we derive a model to evaluate the DR capacity of a building.
In particular, we develop a metric to evaluate the maximum 
reduction of the building's consumption given an occupancy distribution, energy saving features, and external weather conditions.
In our DR model, we focus our attention on a single energy consumer who responds to a single energy provider (or retailer).
We assume that a DR event is successfully cleared if an energy consumer reduces its energy consumption 
by an amount 
that is greater than a pre-agreed quantity. This is done with respect to a baseline consumption, within a pre-determined DR period.

\subsection{Building Energy Consumption and Control Parameters}
Now, we formally describe our model given the aforementioned assumptions.
Let $z'(t)$ denote the total power consumption of a building at time $t$, with upper and lower bounds as follows: $0 \leq z_l \leq z'(t) \leq z_h$.
We use $\alpha$ to denote the pre-determined DR period.
Then the total energy consumption during the DR period $\alpha$ from the DR start time $t_s$ can be expressed by $$z(t_s,\alpha)=\int_{t_s}^{t_s+\alpha}z'(\tau)\,\mathrm{d}\tau.$$ 
We can approximate this energy consumption by $$z(t_s,\alpha) \approx \sum_{t_i \in (t_s,t_s+\alpha) }z'(t_i) \Delta T_s,$$ where $\Delta T_s$ is the sampling rate of power measurements, $z'(t_i)$.
Without loss of generality, let us assume that $\Delta T_s=1sec$
by which $$z(t_s,\alpha)=\sum^{\alpha}_{i,t_i \in (t_s,t_s+\alpha) }z'_i,$$ where $z'_i$ is $i_{th}$ sample collected for $(t_s,t_s+\alpha)$.

Now let us model $z(t_s,\alpha)$ as a random variable in $\mathbb{R}_{[\alpha z_l,\alpha z_h]}$.
We use $\theta$ to denote a set of parameters that affect the total power consumption $z'(t)$ such as
weather conditions, occupancy level or building-specific controls.
Let us assume that the samples $z'_i$  are i.i.d Gaussian random variables for $(t_s,t_s+\alpha)$
such that $$z'_i(\theta) \sim N(\mu_{z'}(\theta), \sigma_{z'}(\theta)),$$ where
$\mu_{z'}(\theta)$  and $\sigma_{z'}(\theta)$ are the expectation and the standard deviation of $z'$
given $\theta$.
Note that, in order to define the Gaussian distribution over a finite real line $z' \in \mathbb{R}_{[z_l,z_h]}$,
we make the mapping of $z'$ such that $(z'<z_l)\mapsto z_l$ and  $(z'>z_h)\mapsto z_h$.
Equivalently, we can model $z'_i$ by $$z'_i(\theta)=\mu_{z'}(\theta) + \varepsilon_{z'}(\theta),$$
where $\varepsilon_{z'}(\theta)$ is a zero-mean i.i.d Gaussian random variable with variance $\sigma^2_{z'}(\theta)$. This variance is the uncertainty in $z'_i(\theta)$ that is captured by $\mu_{z'}(\theta)$.
Now it can be easily seen that $z(t_s,\alpha)$ is a random variable, denoted by $z_{\alpha}(\theta)$
that follows a Gaussian distribution with mean $\alpha \mu_{z'}(\theta)$ and variance $\alpha \sigma_{z'}(\theta)$,
i.e. $$z_{\alpha}(\theta) \sim N(\alpha \mu_{z'}(\theta), \sqrt{\alpha} \sigma_{z'}(\theta)).$$ 
Note that the DR start time $t_s$ in $(t_s,t_s+\alpha)$ is implicitly encoded by $\theta$ in $z_{\alpha}(\theta)$.

Let us define two mutually exclusive sets of parameters such that $\theta =(\theta_r,\theta_c)$,
where $\theta_r$ is a set of \emph{reference parameters} that govern the baseline consumption of a building
and $\theta_c$ is a set of \emph{demand response control parameters} to be modified for DR.
We make the following assumptions for the parameters in developing our model:

\assumption{}\label{assumption:one}
An energy provider and a consumer agree on the amount of energy reduction for a DR event with respect to specific settings of the reference parameter before the DR program starts.

\assumption{}\label{assumption:two}
The building control parameters $\theta_c$ are set to their respective constant default values given $\theta_r$ unless a DR event occurs.

\assumption{}\label{assumption:three}
Changing building control parameters $\theta_c$ will cause a deterministic change in total power consumption without changing $\theta_r$.

\assumption{}\label{assumption:four}
The optimal control policy $\theta^{*}_c$ for a DR event does not change during its DR period. 

We use $\bar{z}_{\alpha}(\theta)$ to denote the expectation of $z_{\alpha}(\theta)$,
or a conditional expectation of $z(t_s,\alpha)$ given $\theta$.
Hence, $z_{\alpha}(\theta_r,\theta_c)$ can be interpreted as the expected baseline energy consumption of a building.
Since $\theta_c$ is deterministic given $\theta_r$ from Assumption~\ref{assumption:two} we will use a shorthand notation, $z_{\alpha}(\theta_r)$ instead of  $z_{\alpha}(\theta_r,\theta_c)$ when it is clear from context.

\subsection{Demand Response Capacity Formulation}
Let $u'(t)$ denote a utility function 
that quantifies the occupant comfort level at time $t$ as a real number between $0$ and $1$.
Then let us define $u(t_s,\alpha)=\frac{1}{\alpha}\int_{t_s}^{t_s+\alpha}u'(\tau)\,\mathrm{d}\tau$,
i.e. the average utility during the DR period $(t_s,t_s+\alpha)$.
Then, similar to $z_{\alpha}(\theta)$, we define a conditional random variable $u_{\alpha}(\theta) \in \mathbb{R}_{[0,1]}$ for $u(t_s,\alpha)$,
and its expectation $ \bar{u}_{\alpha}(\theta)=E[u(t_s,\alpha)|\theta]$ .
Let $\theta_c^{*}$ denote the optimal control parameters for the optimization problem in (\ref{eq:optimal_params}),
\begin{equation}
\begin{aligned}
& \theta_c^{*}=\underset{\theta_c}{\text{argmin}}
& & \bar{z}_{\alpha}(\theta_r,\theta_c) \\
& \text{subject to}
& & \bar{u}_{\alpha}(\theta_r,\theta_c) \geq u_{min}(\theta_r)\\
\end{aligned}
\label{eq:optimal_params}
\end{equation}
where $u_{min}(\theta_r)$ is the minimum utility value required for occupants in the building, which depends on $\theta_r$.
We define the DR capacity of a building $C_{\alpha}(\epsilon)$ as follows: 
\begin{equation}\label{eqn:capacity_model}
C_{\alpha}(\epsilon)=\underset{\beta}{\text{argmax}}\left\{~Pr(\bar{z}_{\alpha}(\theta_r)-z_{\alpha}(\theta_r,\theta_c^{*}) \leq \beta) \leq \epsilon \right\}
\end{equation}
where $\beta \geq 0$ is the amount of energy reduction requested by an energy provider during the DR period $\alpha$.
We consider $\beta=0$ as a special case where an energy provider requests \emph{zero} energy reduction during $\alpha$, i.e.
request to maintain the expected energy consumption. We can interpret $C_{\alpha}(\epsilon)$ as the maximally \emph{pre-agreed} energy reduction capacity for DR given $\epsilon$, which is a tolerance limit for the uncertainty associated with successfully clearing a DR request.

Let us derive a formula for the probability of successfully clearing a demand response event, as a function of the requested energy reduction $\beta$: 
\begin{equation}
f_\theta (\beta)=Pr(\bar{z}_{\alpha}(\theta_r)-z_{\alpha}(\theta_r,\theta_c^{*})\leq \beta).
\end{equation}
We define the amount of energy reduction by demand response control $\theta_c^{*}$ as
\begin{equation}
\Delta z_{\alpha}(\theta_r,\theta_c^{*})=z_{\alpha}(\theta_r,\theta_c)-z_{\alpha}(\theta_r,\theta_c^{*}),
\end{equation}
 where $\Delta z_{\alpha}(\theta_r,\theta_c^{*}) \geq 0$. 
From Assumption~\ref{assumption:three}, $\Delta z_{\alpha}(\theta_r,\theta_c^{*})$ is a deterministic quantity given $\theta_c^{*}$ as opposed to a random variable $z_{\alpha}(\theta_r)$. Hence, it represents the intrinsic curtailment capability of the building's controllable loads given $\theta_r$. We will use a shorthand notation, $\Delta z_{\alpha}(\theta_c^{*})$ instead of  $\Delta z_{\alpha}(\theta_r,\theta_c^{*})$.

By rearranging terms we have
\begin{align}\label{eqn:dervQfunc}
 f_\theta (\beta) &= Pr\left(\bar{z}_{\alpha}(\theta_r)+\Delta z_{\alpha}(\theta_c^{*})- \beta \leq z_{\alpha}(\theta_r)\right)  \nonumber \\
 &\overset{(*)}{=} Q\left(\frac{\bar{z}_{\alpha}(\theta_r)+\Delta z_{\alpha}(\theta_c^{*})-\beta-E[z_{\alpha}(\theta_r)]}{\sqrt{Var(z_{\alpha}(\theta_r))}}\right)  \nonumber  \\
 &\overset{(**)}{=} Q\left(\frac{\Delta z_{\alpha}(\theta_c^{*})-\beta}{\sqrt{\alpha}\sigma_{z'}(\theta_r)}\right)
\end{align}
where $Q(x)$ is the Q-function: the tail probability of the standard normal distribution $N(x|0,1)$, which can be
 defined by $Q(x)=1-\int_{-\infty}^{x}N(x'|0,1)\,\mathrm{d}x'$.
The equality (*) in (\ref{eqn:dervQfunc}) is derived by expressing $P(x\leq z_{\alpha}(\theta_r))$ by $Q(x)$.
The equality (**) holds by the definition  $\bar{z}_{\alpha}(\theta_r)=E[z_{\alpha}(\theta_r)]$.

\begin{figure}[t]
  \centering
 \includegraphics[width=\linewidth]{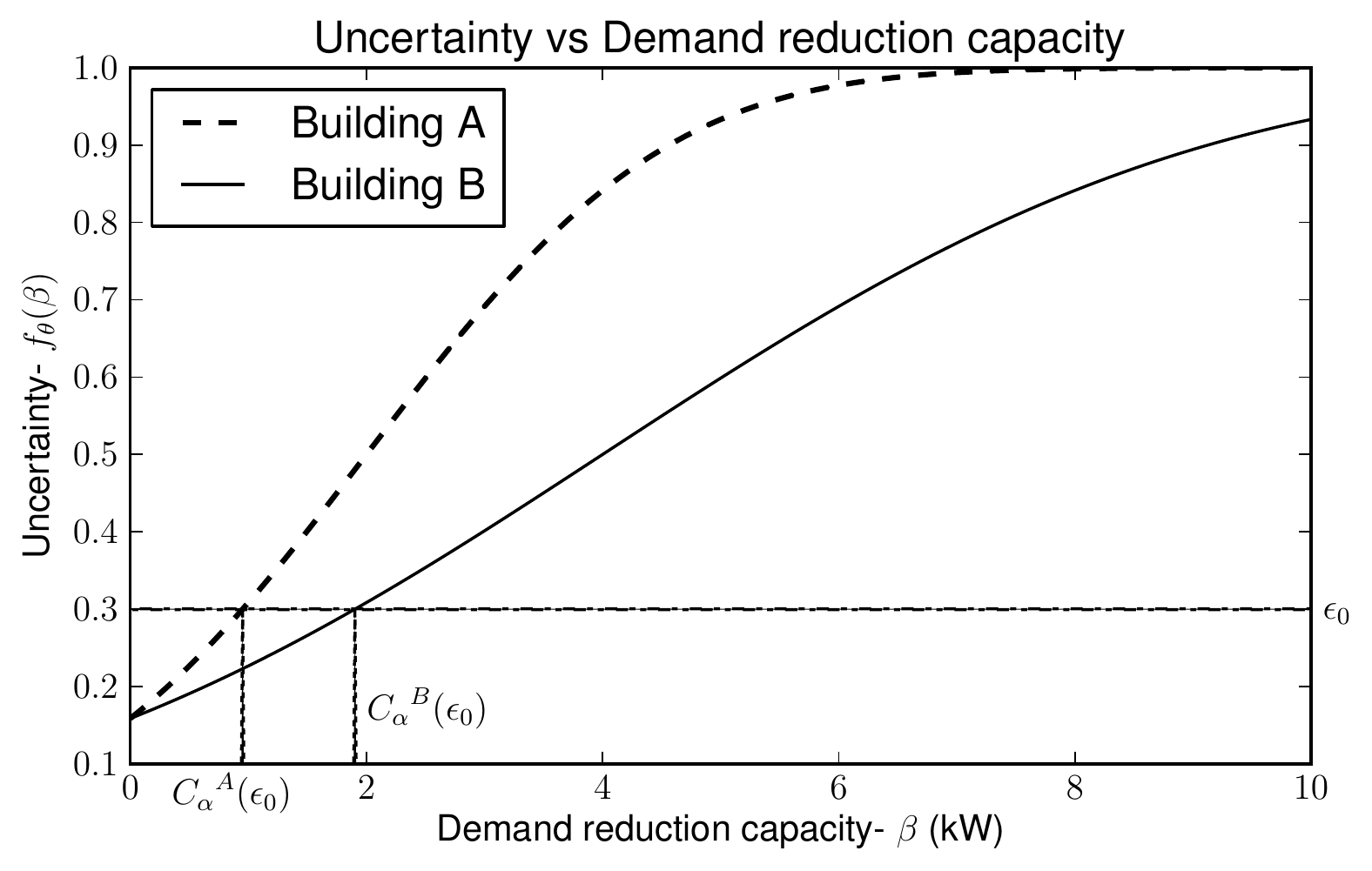}
  \caption{Illustrative example of relationship between DR capacity and uncertainty. Building B exhibits more ideal behavior compared to Building A as an increase in capacity leads to a lesser increase in uncertainty.}
  \label{fig:illustration}
\end{figure}

If $\Delta z_{\alpha}(\theta_c^{*}) \geq 0$, $f_\theta (\beta)$ is an increasing function of $\beta$ and the building's demand response capacity is determined by the uncertainty threshold value $\epsilon$. Each building will exhibit a unique trade-off between DR capacity and the uncertainty of realizing that curtailment. This is illustrated in Figure~\ref{fig:illustration} for two hypothetical buildings (not the testbeds) and an uncertainty threshold of $\epsilon=0.3$. 

The final formula for $f_\theta (\beta)$ in (\ref{eqn:dervQfunc}) is analogous to the Bit Error Rate (BER): a fundamental performance metric in communication theory~\cite{proakis2001}. The BER is the number of bit errors divided by the total number of transferred bits during a studied time interval, and it is often expressed as a Q-function of the Signal-to-Noise Ratio. In 
 (\ref{eqn:dervQfunc}), we draw the following parallels:
\begin{itemize}
\item $\sqrt{\alpha}\sigma_{z'}(\theta_r)$ can be interpreted as the additive white Gaussian noise (AWGN) of a communication channel (background power fluctuation without DR)
\item $\Delta z_{\alpha}(\theta_c^{*})$ can be interpreted as the transmitted signal strength (energy reduction when loads are controlled during DR)
\item $\beta$ in $f_\theta (\beta)$ can be interpreted as the decision threshold for a bit (pre-agreed energy reduction for a successful DR event)
\end{itemize}


The formula for $f_\theta (\beta)$ also reveals some interesting insights into the interactions between $\beta$,
$\alpha$, $\sigma_{z'}(\theta_r)$, and $\Delta z_{\alpha}(\theta_c^{*})$.
First, the uncertainty $f_\theta (\beta)$ will increase for higher $\sigma_{z'}(\theta_r)$ if $\beta < \Delta z_{\alpha}(\theta_c^{*})$, 
where the amount of requested energy reduction for DR is within the building's intrinsic curtailment capability. 
In this case, $\sigma_{z'}(\theta_r)$ represents noise.
However, the opposite occurs if $\beta > \Delta z_{\alpha}(\theta_c^{*})$, where the requested energy reduction is higher than the building's capacity (i.e. over-committed DR). Higher background power fluctuation opportunistically helps to meet the over-committed DR. Note that we have $f_\theta (\beta)\geq 0.5$ if $\beta > \Delta z_{\alpha}(\theta_c^{*})$
which should not be considered for any DR program.

Given that $Q(-x)$ is an increasing function of $x$, the final exact solution of the demand response capacity $C_{\alpha}(\epsilon)$ in (\ref{eqn:capacity_model}) can be simply found by its boundary value at $\epsilon$ as follows: 
\begin{empheq}[box=\fbox]{align}
  C_{\alpha}(\epsilon)=\Delta z_{\alpha}(\theta_c^{*}) -\sqrt{\alpha}\sigma_{z'}(\theta_r) Q^{-1}(\epsilon))
 \label{eqn:capa_final}
\end{empheq}
where  $Q^{-1}(x)$ represents the inverse function of $Q(x)$.
It is easily seen that $C_{\alpha}(\epsilon)$ is an increasing function of $\epsilon$
and the rate of capacity reduction for $\epsilon$ given DR period $\alpha$ is determined by $\sqrt{\alpha}\sigma_{z'}(\theta_r)$.

\begin{table}
\centering
\scriptsize
\caption{\footnotesize Classification of reference parameters for demand response event}\label{tab:paramQt}
\begin{tabular}{c|c|c|c|c|c|c}
\hline
State          & 0      & 1       & 2      &   3    &  4       &   5     \\ \hline
\multirow{2}{*}{$\theta^{wd}_r$}& Sat,Sun     &  Mon $\sim$ & \multirow{2}{*}{-} &  \multirow{2}{*}{-} & \multirow{2}{*}{-} & \multirow{2}{*}{-} \\
                                & ,Pub.Hd      & Friday &       &       &    &        \\
$\theta^{hr}_r$ & 21-7 & 7-10   & 10-12 & 12-14 & 14-18  &  18-21 \\
$\theta^{occ}_r$& 0 & 0-25 & 25-50 & 50-75 & $>$75  & - \\
$\theta^{sol}_r$ & 0 & 0-200 & 200-400 & 400-600 & $>$600 &    - \\
$\theta^{temp}_r$ & $<$21 & 21-24 & 24-27 & 27-30 & $>$30 &    -   \\

\hline
 \end{tabular} \\
{\footnotesize Note: the units for ($\theta^{hr}_r$, $\theta^{occ}_r$, $\theta^{sol}_r$,$\theta^{temp}_r$) are (hrs, ($\%$), ($W/m^2$), $\celsius$)}
\end{table}

\begin{table}
\centering
\scriptsize
\caption{\footnotesize Lookup table example for parameters and total consumption statistics without DR}\label{tab:lookup}
\begin{tabular}{c|c|c|c|c}
  \hline
 $(\theta^{wd}_r ,\theta^{hr}_r,\theta^{occ}_r,\theta^{sol}_r,\theta^{temp}_r)$ & $\theta^{light}_c$ & $\theta^{hvac}_c$
 & $\bar{Z'}(\theta)$ & $\sigma_{z'}(\theta)$ \\
   \hline
  $(1 , 4 , 2 , 2 , 3)$ & ON & $23\celsius$ & 1.2 kW & 0.2 kW\\
 $\vdots$ & $\vdots$ &$\vdots$ & $\vdots$&$\vdots$\\
  \hline
 \end{tabular} \\
\end{table}

\section{Demand Response Capacity Analysis} \label{sec:analysis}
In this section, we apply the DR capacity model derived in Section \ref{sec:capacity_model} to
the datasets from the BCA and ADSC testbed described in Section \ref{sec:testbeds}.
We construct a look-up table from each testbed site's sensor dataset to search the parameters as well as other quantities to estimate DR capacity.

\subsection{Parameters}
For the reference parameters $\theta_r$, we consider 
working or non-working days ($\theta^{wd}_r$), hours of the day ($\theta^{hr}_r$), occupancy level ($\theta^{occ}_r$), external solar irradiance ($\theta^{sol}_r$), and external temperature ($\theta^{temp}_r$). 
Non-working days include Saturday, Sunday, and public holidays. 
Non-working days include Saturday, Sunday, and public holidays. 
We assume that the energy consumer and the provider agree to the classification of
the state spaces of $\theta_r$ to evaluate a baseline consumption $z_{\alpha} (\theta_r)$ before the demand response program starts.
Hence, well-classified states of reference parameters allow them to reduce uncertainty in demand response.
We use the same classification of $\theta_r$ for both ADSC and BCA testbed to make fair comparisons between them in demand response capacity. Table \ref{tab:paramQt} explains the details of the classification.

For control parameters $\theta_c$, we consider only two types of loads: HVAC and lighting. These are denoted by $\theta^{hvac}_c$ and $\theta^{light}_c$, respectively.
As opposed to reference parameters, the control parameters are defined separately for ADSC and BCA in order to take account of their particular control capabilities.
Note that we assume plug loads are not controllable for DR. This is generally true for our testbeds as most plug loads are computers.
Hence, a large variance in plug load consumption will cause a higher $\sigma_{z'}(\theta)$.
For $\theta^{hvac}_c$, we consider a temperature set-point control for ADSC and an ON/OFF control of the ventilation fans for the ZEB. 
For $\theta^{light}_c$, we consider an ON/OFF control for ADSC and a dimmer switch for ZEB.

Next, we build and update a look-up table that stores the default control parameters ($\theta^{light}_c,\theta^{hvac}_c$) and the total power consumption statistics ($\bar{Z'}(\theta), \sigma_{z'}(\theta)$) for each instance of an observed state of reference parameters $(\theta^{wd}_r ,\theta^{hr}_r,\theta^{occ}_r,\theta^{sol}_r,\theta^{temp}_r)$ as shown in Table \ref{tab:lookup}.
The control parameter states of each row in the table represent the default control setting, without a DR event, under the conditions given by the reference parameters. 

\begin{figure}
  \centering
 \includegraphics[width=\linewidth]{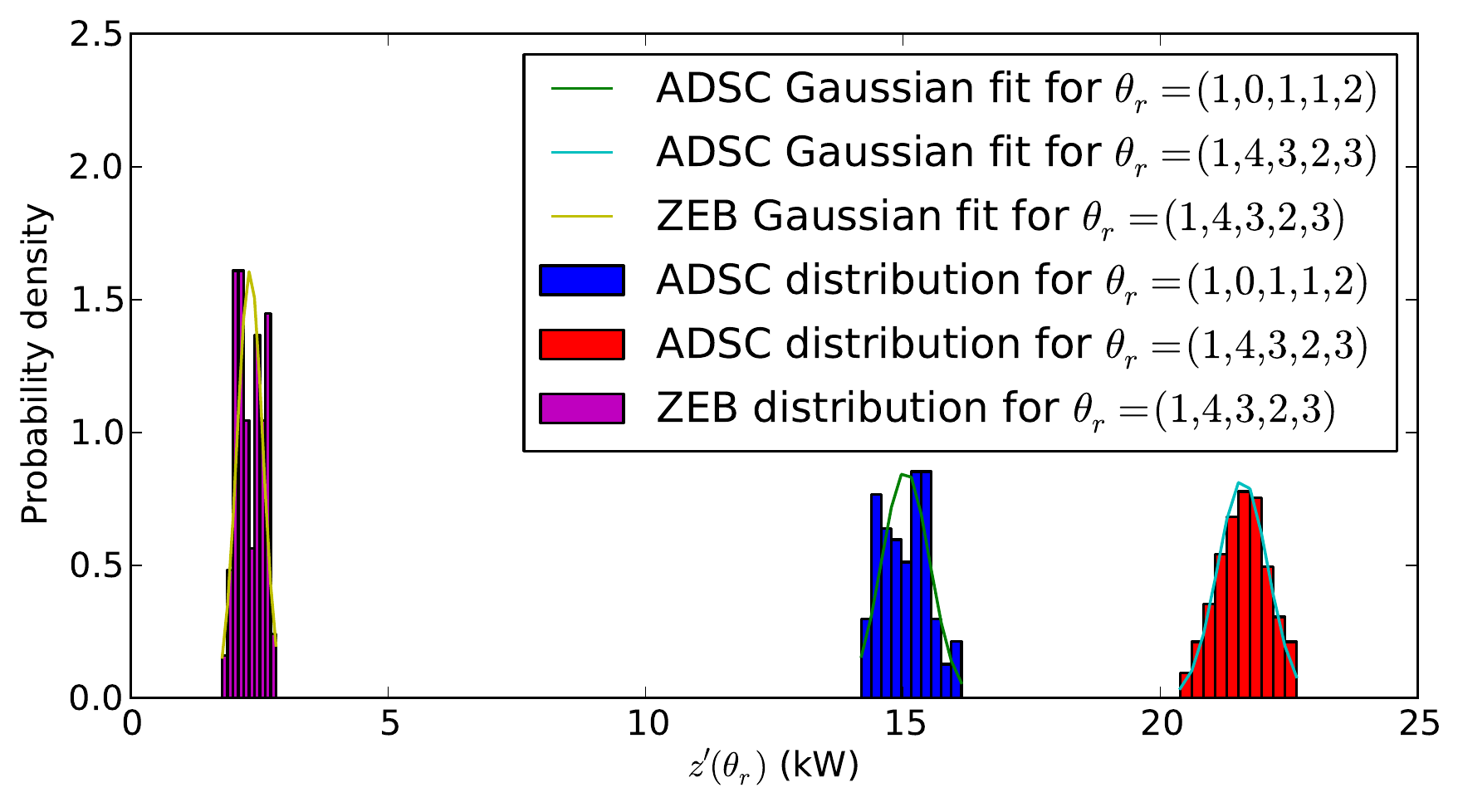}
  \caption{Conditional distributions of $z'_{\alpha} (\theta_r)$ for different values of $\theta_r = (\theta^{wd}_r,\theta^{hr}_r,\theta^{occ}_r,\theta^{sol}_r,\theta^{temp}_r)$ in ADSC and BCA ZEB testbeds . Note the parameters are the column indices that map to values given in Table \ref{tab:paramQt}.}
  \label{fig:joint_dist}
\end{figure}

Figure \ref{fig:joint_dist} shows examples of the conditional distribution of $z'(\theta_r)$ for specific values of $\theta_r$ from Table \ref{tab:lookup}.
In the figure, the distribution of total power consumption samples given $\theta_r$ can be well approximated by a Gaussian distribution, validating our assumption $z'_i(\theta) \sim N(\mu_{z'}(\theta), \sigma_{z'}(\theta))$.

\subsection{Demand Response Controls}
\begin{table}
\centering
\scriptsize
\caption{\footnotesize Minimum utility required given occupancy states}\label{tab:util}
\begin{tabular}{c|c|c|c|c|c|c}
\hline
 \multicolumn{2}{c|}{$\theta^{occ}_{r}$} &  0       & 1           & 2       &   3      & 4    \\ \hhline{=|=|=|=|=|=|=}
Thermal  &$u_{min}$                       & 0.1      &  0.5        & 0.7     &  0.8     & 0.9       \\ \cline{2-7}
 Comfort  & $\celsius$                   & 30        &  27          &  27      &  26       & 26      \\ \hline
ADSC      &$\theta^{*}_c$               & 30        &  27          &  27      &  26       & 26       \\ \cline{2-7}
          &$z'(\theta^{*}_c) (kW)$     & 0        &  0      &  0    &  2.778    & 2.778      \\ \hline
ZEB       &$\theta^{*}_c$                & OFF        &  OFF       &  OFF      &  ON     & ON       \\ \cline{2-7}
          &$z'(\theta^{*}_c) (kW)$     & 0        &  0          &  0      &  1.5       & 1.5      \\ \hhline{=|=|=|=|=|=|=}
Visual    &$u_{min}$                      & 0.1      &  0.5        & 0.7     &  0.8     & 0.9    \\ \cline{2-7}
Comfort   &$Lux$                          & 0        &  500          &  500      &  600       & 600   \\  \hline
ADSC      &$\theta^{*}_c$                & OFF        &  OFF          &  OFF      &  OFF     & OFF      \\ \cline{2-7}
          &$z'(\theta^{*}_c) (kW)$    & 0        &  0          &  0      &  0       & 0       \\ \hline
ZEB       &$\theta^{*}_c$                & OFF        &  OFF          &  OFF      &  600       & 600      \\ \cline{2-7}
          &$z'(\theta^{*}_c) (kW)$     &    0     &  0       &  0     &  0.136      & 0.136      \\ \hline
 \end{tabular}\\
{\footnotesize  Note: values based on ($\theta^{wd}_r$,$\theta^{hr}_r$, $\theta^{sol}_r$,$\theta^{temp}_r$) = (1,4,2,3), $\alpha=1hr$ }
\end{table}

It is often impractical to solve the optimization problem in (\ref{eq:optimal_params}) since generally we do not have an accurate model for the objective $\bar{z}_{\alpha}(\theta_r,\theta_c)$ and the constraint function $\bar{u}_{\alpha}(\theta_r,\theta_c)$).
Instead, we build a look-up table to search $\theta_c^{*}$ for lighting and HVAC.
Table \ref{tab:util} shows an example of such look-up table for ZEB and ADSC.
In the table, we use indoor temperature/humidity and lux measurements to evaluate visual and thermal comfort level which are mapped into a 0 to 1 scale by their respective utility functions. More details about the utility functions for visual and thermal comfort can be found in our previous work \cite{jung2013energytrack}. Then, $\Delta z_{\alpha}(\theta_c^{*})$ is found by cross-referencing Table \ref{tab:util} and Table \ref{tab:lookup} for $\theta_r$.

In our analysis we assume that the minimum utility requirement is only dependent on occupancy level, i.e. $u_{min}(\theta^{occ}_{r})$.
The optimal parameter $\theta_c^{*}$ is determined by finding the setting such that
the utility is maintained above $u_{min}(\theta^{occ}_{r})$ with the lowest energy consumption during the demand response period $\alpha$.
Intuitively, $\theta^{*}_c$ for HVAC is heavily dependent on a building's heating inertia as better thermal insulation 
can allow buildings to maintain a desired indoor temperature longer after the HVAC is turned off.
Similarly, the design of a building for natural lighting will greatly impact on $\theta^{*}_c$ for lighting since daylight penetrating inside building can be sufficient to support the minimum lux requirement without internal lighting.
For both HVAC and lighting, a longer demand response period $\alpha$ means fewer degrees of freedom are available for the optimization of $\theta_c$, which results in smaller $\Delta z_{\alpha}(\theta_c^{*})$.

\subsection{Evaluation of Demand Response Capacity}

\begin{figure}[t]
  \begin{center}
    \begin{tabular}{c}
 \includegraphics[width=0.9\linewidth]{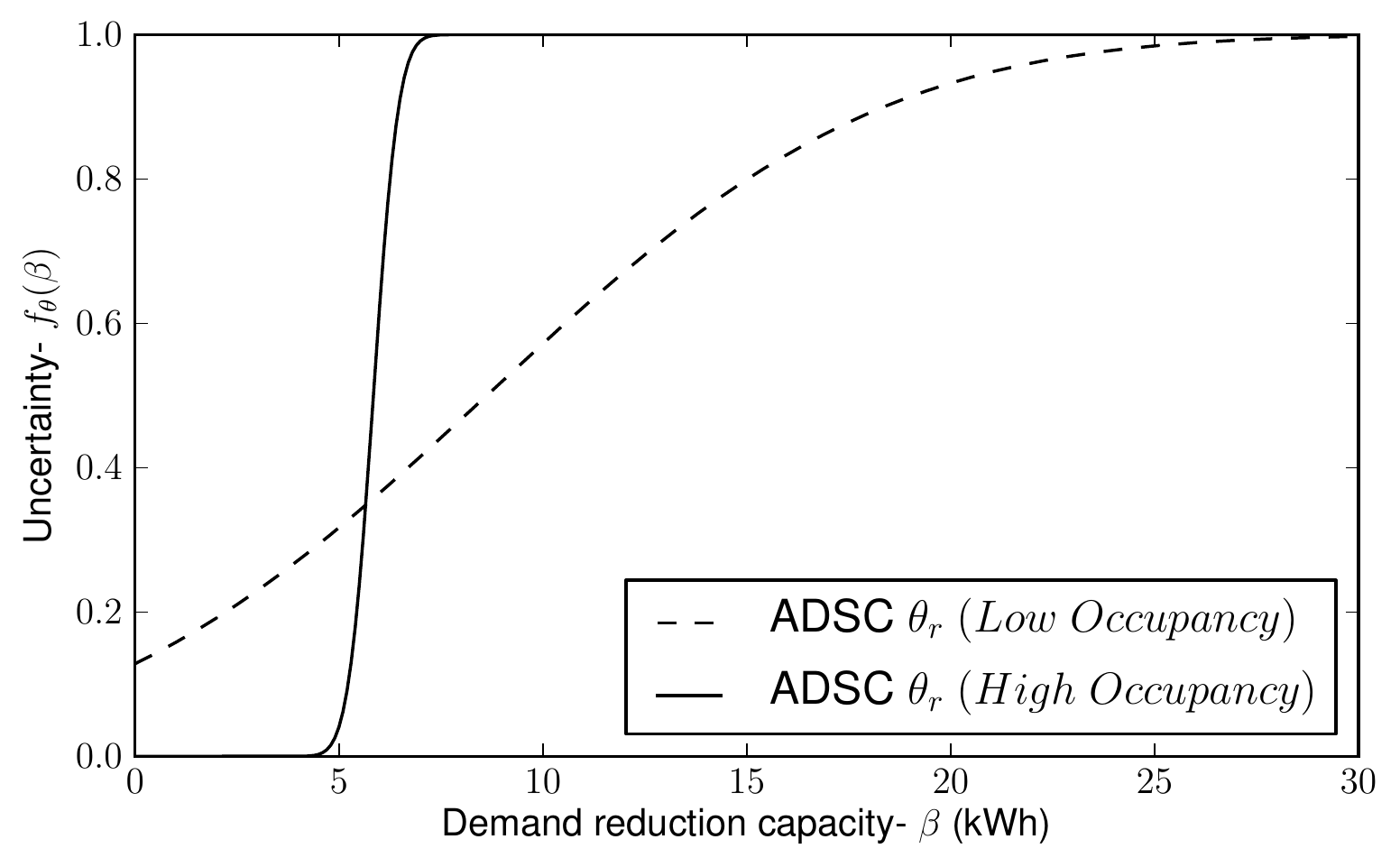}\\ {\small (a) ADSC testbed} \\
  \includegraphics[width=0.9\linewidth]{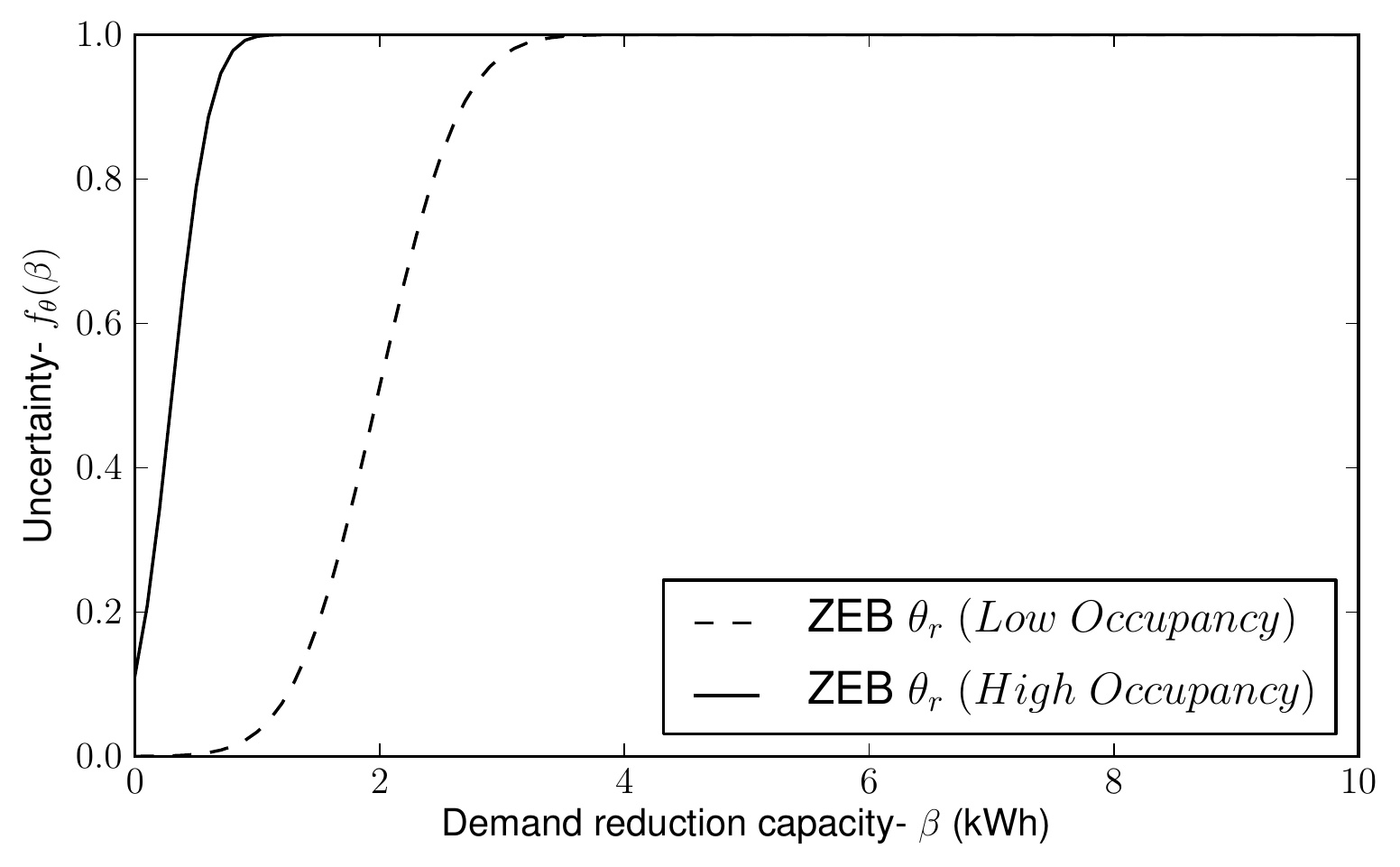}\\ {\small (b) ZEB testbed} \\
    \end{tabular}
  \caption{\footnotesize Uncertainty-DR capacity trade-off comparison for $\alpha=1hr$, where 
  $\theta_r = (1, 4, \theta^{occ}_r, 2, 3)$ and $\theta^{occ}_r = 1$ (low) or $3$ (high).}
\label{fig:capa_cmpr}
  \end{center}
\end{figure}
In Figure \ref{fig:capa_cmpr} we evaluate and compare the DR capacity of the ADSC and ZEB testbeds for different
reference parameter settings $\theta_r$, with a demand response period of $\alpha=1hr$.
For each testbed we compute the uncertainty-DR capacity trade-off for $\theta^{occ}_r=1,3$ (i.e. low and high occupancy states)
while the rest of reference parameters are fixed at $\theta^{wd}_r$=1 (i.e. working day), $\theta^{hr}_r$=4 (i.e. 2-6pm), $\theta^{sol}_r$=1 (i.e. sunny to partly cloudy), and $\theta^{temp}_r$=3 (i.e. warm weather).

For the ZEB, the DR capacity at the low-occupancy state is higher than at the high-occupancy state for all uncertainty values. For a given uncertainty of $\epsilon=0.2$ the gain in DR capacity from low occupancy levels to high for ZEB  is 1.438kWh compared to -3.212kWh for ADSC.
The difference in DR capacity gain can be explained by the fact there is significantly greater variation in the loads at ADSC for low occupancy levels leading to a high $\sigma_{z'}(\theta_r)$. This causes an increase in the elasticity of capacity in response to changes in uncertainty requirements, hence the slope is more gradual. The great variation arises because low occupancy levels are rarely seen during the 2-6pm period at ADSC and occupants may or may not turn off their plug loads when they are not around.

\begin{figure}[t]
  \centering
 \includegraphics[width=0.9\linewidth]{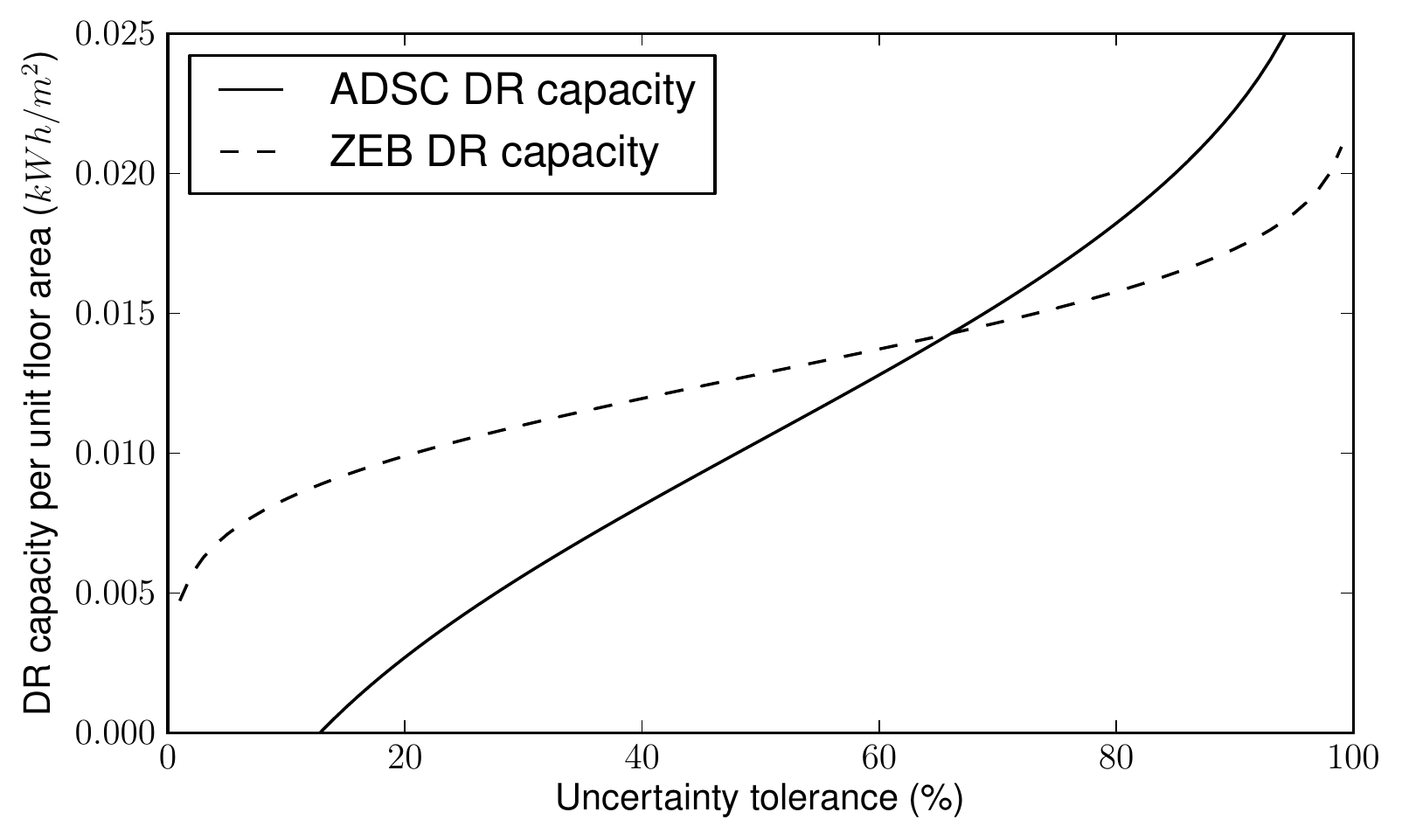}
  \caption{DR capacity comparison between ADSC and ZEB at low occupancy given $\theta_r = (1,4,1,2,3)$ and $\alpha=1hr$.}
  \label{fig:capa_cmpr_all}
\end{figure}
In Figure \ref{fig:capa_cmpr_all} we compare the DR capacity $C_{\alpha=1hr}(\epsilon)$ for ADSC and the ZEB after normalizing the values by their floor area.
It is clear from the figure that the ZEB has greater DR capacity at low and moderate uncertainty tolerance values ($< 60\%$). This is likely due to the presence of more precise control capabilities, i.e. dimmable lighting (see Figure~\ref{fig:avgLightingPwr}). 
When the uncertainty tolerance exceeds $60\%$, ADSC does have a larger DR capacity than the ZEB; however, such high uncertainty about DR performance would likely be undesirable for building owners and/or aggregators. This is particularly true in market settings that enforce non-compliance penalties or compensate demand resources based on the ratio of actual and expected response during events~\cite{singaporeIL}.

\begin{figure}[t] 
  \centering
 \includegraphics[width=0.85\linewidth]{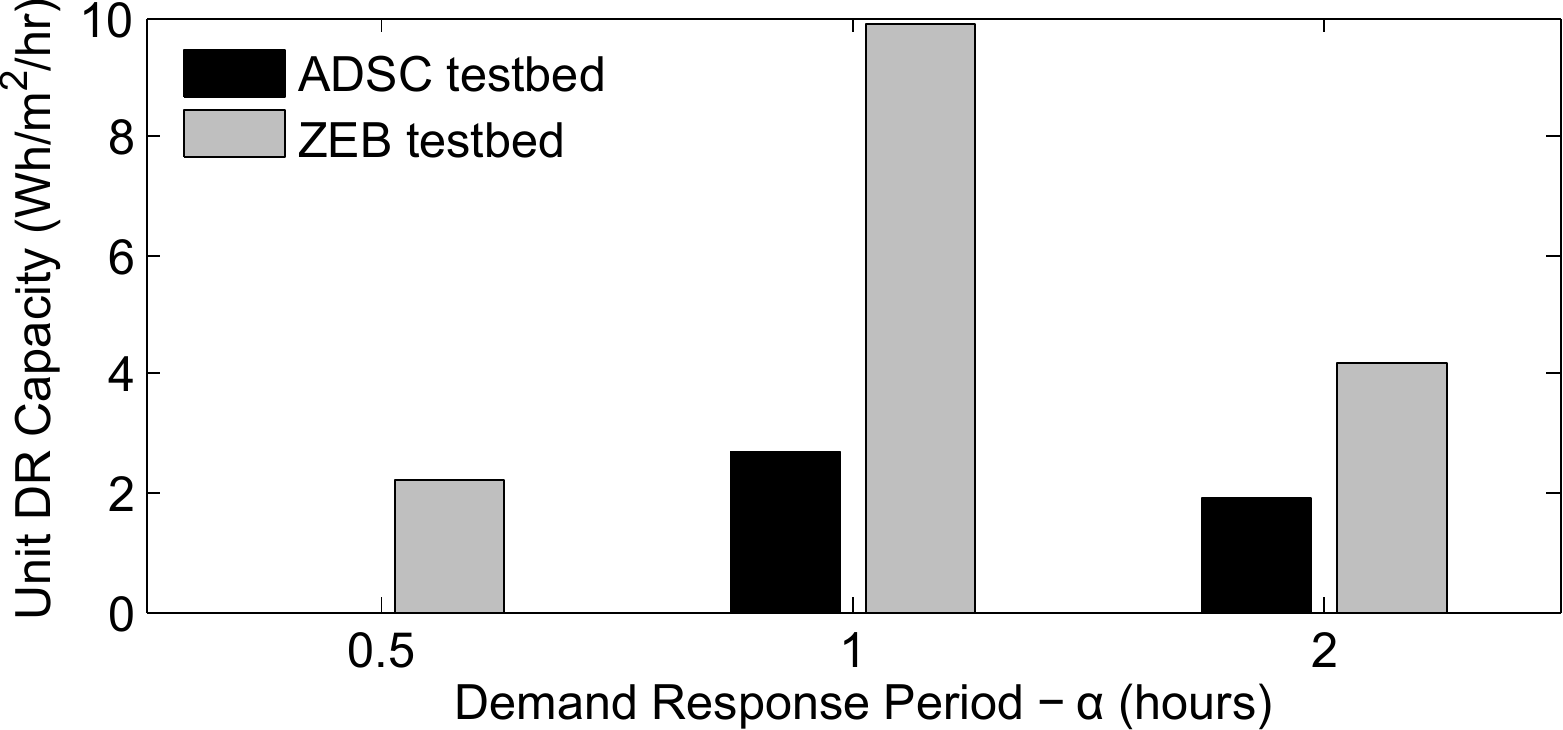}
  \caption{DR capacity comparison for different demand response periods $\alpha$ given $\theta_r=(1,4,1,2,3)$.}
  \label{fig:DRcapacity_compare}
\end{figure}

Finally, we compare demand response capacity $C_{\alpha}(\epsilon=0.1)$ for different DR periods $\alpha$ = $0.5hr$, $1hr$, and $2hr$ for ADSC and the ZEB testbeds in Figure~\ref{fig:DRcapacity_compare}. 
We normalize DR capacity by the testbeds' respective floor areas for each value of $\alpha$ for a fair comparison. We refer to this as \emph{unit DR capacity}. The figure shows that both the ZEB and ADSC have the highest unit DR capacity at $\alpha=1hr$. In particular, for ADSC no DR capacity is available if the DR period is set to $0.5$ hours because the variance of uncontrollable loads (i.e. plug loads) during the DR period always causes a larger uncertainty than the tolerance limit of $\epsilon=0.1$. This implies that the demand response period must be carefully chosen with respect to the building's particular reference parameters.





\section{Conclusion and Future Work}
\label{sec:conclude}
In this paper, we have presented a framework to evaluate the demand response capacity for buildings. We deployed sensor networks in two testbeds, collected data from the sensors over a period of 10 weeks and used our framework to evaluate the DR capacity for these testbeds. In future work, we will extended our model to more complex scenarios that may contain multiple DR participants and DR aggregators.

\section*{Acknowledgments}
This work was supported by Singapore's Agency for Science, Technology, and Research (A*STAR), through a research grant for the Human Sixth Sense Programme at the Advanced Digital Sciences Center, and by the Korea Micro Energy Grid (KMEG) of the Office of Strategic R\&D Planning (OSP), Korea government Ministry of Knowledge Economy (No. 2011T100100024). 
We thank the Building and Contruction Authority of Singapore (BCA) for providing a testbed environment and permitting the use of project data for this publication. We would also like to thank the team at the Nanyang Technological University Intelligent Systems Centre (IntelliSys) for their help with sensor network deployment and system integration.
%

\balance

\bibliographystyle{IEEEtran}
\bibliography{adsc_smartgridcomm-final}
\end{document}